\documentclass[journal,doublecolumn]{IEEEtran}
\usepackage{amssymb}
\usepackage{cite}

\ifCLASSINFOpdf
\else
\fi
\usepackage{amsfonts}
\usepackage{amsmath}
\usepackage{latexsym,bm}
\usepackage{amsthm}
\newtheorem{theorem}{Theorem}
\newtheorem{corollary}{Corollary}
\newtheorem{remark}{Remark}
\usepackage{array}
\usepackage{fixltx2e}
\usepackage{graphicx}
\usepackage{epstopdf}
\usepackage{float}
\usepackage{xcolor}
\usepackage[numbers]{natbib}
\usepackage{subfigure}

\begin{document}
\title{Hardware Impairments Aware Full-Duplex NOMA Networks Over Rician Fading Channels}
\author{Author A, Author B, Author C, Author D}
\author{Chao Deng, Meng~Liu, \emph{IEEE}, \emph{Student Member}, Xingwang Li, \emph{IEEE}, \emph{Senior Member}, Yuanwei Liu, \emph{IEEE}, \emph{Senior Member}
\thanks{C. Deng, M. Liu and X. Li are with the School of Physics and Electronic Information Engineering, Henan Polytechnic University, China (email:lixingwangbupt@gmail.com).}
\thanks{Y. Liu is with the School of Electronic Engineering and Computer Science, Queen Mary University of London, UK (email:yuanwei.liu@qmul.ac.uk).}}
\maketitle
\begin{abstract}
A cooperative full duplex (FD) non-orthogonal multiple access (NOMA) scheme over Rician fading channels is considered. To be practical, imperfect successive interference cancellation (ipSIC) and residual hardware impairments (RHIs) at transceivers are taken into account. To evaluate the performance of the considered system, the analytical approximate expressions for the outage probability (OP) and the ergodic rate (ER) of the considered system are derived, and the asymptotic performance is also explored. Simulation results manifest that the FD cooperative NOMA can improve the ergodic sum rate (ESR) of the system compared to the half duplex (HD) mode.
\end{abstract}


\section{Introduction}
\IEEEPARstart{T}{he} fifth generation (5G) cannot only greatly improve the high broadband service experience of mobile internet users, but also fit the needs of large connectivity and wide coverage of mobile internet services \cite{liu217}. As a key technology of 5G, non-orthogonal multiple access (NOMA) can improve spectrum efficiency, reduce the power consumption of equipments and network transmission time delay, and improve the reliability of network transmission, which has been accepted by the Third Generation Partnership Project (3GPP) \cite{ding2015}. The main advantage of NOMA technology is that the transmitter introduces the interference signals actively, and then uses the successive interference cancellation (SIC) technology to realize the correct demodulation at the receiver, which may lead to a certain degree of complexity of the receiver, so that it can improve the utilization of spectrum resources greatly \cite{dail}.

Cooperative communication is introduced into NOMA due to its higher diversity gain and extensive coverage. In the past, many related works have only focused on half-duplex (HD) relays cooperative communication systems, e.g. see \cite{ding2015} and \cite{lixingwsj}. In order to reduce the waste of spectrum resources caused by this situation, the full-duplex (FD) relay is taken into account in the cooperative NOMA systems. Unlike HD mode, FD can receive and transmit signals in the same time slot, which increases the transmission rate of the system to a great extent. In \cite{ding2018}, authors have proven the outage probability (OP) that the FD mode coexists with NOMA, which paves the way for the subsequent derivation of the letter. The performance of the FD cooperative NOMA system over Nakagami-$m$ fading channels was discussed in \cite{zhong2016}. However, previous work was focus on the ideal hardware conditions. In practice, due to incomplete matching of components or manufacturing process problems, residual hardware impairments (RHIs) will inevitably occur, hence affecting the performance of radio frequency (RF) transceivers \cite{lixingwsj}. Furthermore, considering the high complexity of receiver processing signals and error propagation, we integrate imperfect SIC (ipSIC) and RHIs into the system.

Motivated by the above discussion, in this letter, this letter investigates the performance of FD cooperative NOMA system with RHIs and ipSIC over Rician fading channels. In order to evaluate the performance of the system, the analytical expressions of the OP and the ergodic sum rate (ESR) in the case of high signal-to-noise ratio (SNR) are derived. The simulation results show that RHIs and ipSIC have a negative impact on the system performance. More specifically, the impact of ipSIC on the system is more pronounced than that of RHIs.

\section{System Model}
\subsection{Information Transmission}
\begin{figure}[!tp]
\setlength{\abovecaptionskip}{0pt}
\centering
\includegraphics [width=2.3in]{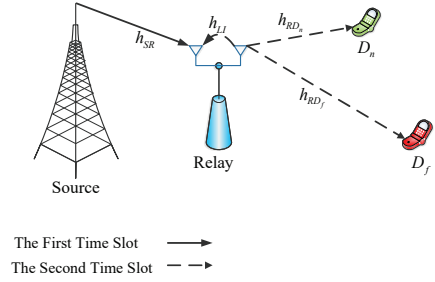}
\caption{The diagram of full duplex cooperative NOMA system}
\label{fig1}
\end{figure}

As shown in Fig. 1, we consider a dual-hop NOMA FD cooperative relay system, which consists of one source $S$, one FD relay $R$, and two destination nodes $D_f$ (the far user) and $D_n$ (the near user). We assume that both source and users are equipped with a single antenna, while the relay is equipped with a receive antenna and a transmit antenna. We also assume that all the channels are Rician channels and the direct links between the source and the users are inexistent due to the the physical obstacles or poor channel conditions, especially in some urban areas \cite{lixingwsj}. Thus, the direct links between the source and the users are not considered in this paper.

During the $t$-th time slot, $S$ sends composite signal ${y_{{S_t}}} = \sqrt {{a_1}{{ P}_S}} {x_1}\left[ t \right] + \sqrt {{a_2}{{ P}_S}} {x_2}\left[ t \right]$ to $R$, where ${{ P}_S}$ is the total transmitted power of $S$; $x_1$ and $x_2$ are corresponding signals of $D_f$ and $D_n$ with $\mathbb{E}\left[ {{{\left| {{x_1}} \right|}^2}} \right] = \mathbb{E}\left[ {{{\left| {{x_2}} \right|}^2}} \right] = 1$, where $\mathbb{E}\left[ X \right]$ denotes the expectation operator of random variable $X$; $a_1$ and $a_2$ are the power allocation coefficients of $S$ transmitted signals with ${a_1} + {a_2} = 1$ and ${a_1} > {a_2}$. Thus, the received signal at $R$ can be expressed as
\begin{equation}
\label{yr}
{y_R} = {h_{SR}}\left( {{y_{{S_t}}} + {\eta _1}} \right) + {h_{LI}}\varpi \left( {{y_{LI}} + {\eta _2}} \right) + {n_0},
\end{equation}
where ${h_{SR}}$ and ${h_{LI}}$ denote the channel coefficients of $S \to R$ and $R \to R$ links, ${y_{LI}} = \sqrt {{{ P}_R}} {x_{LI}}\left[ {t - \tau } \right]$, ${{ P}_R}$ is the transmitted power of $R$, ${x_{LI}}$ is the signal of the loop self-interference (LSI) at the FD relay with $\mathbb{E}\left[ {{{\left| {{x_{LI}}} \right|}^2}} \right] = 1$ and $\tau $ is the time delay of information transmission; $\varpi $ represents the conversion factor of the relay working mode, $\varpi  = 0$ and $\varpi  = 1$ indicate that the relay is in HD mode and FD mode, respectively; ${\eta _1} \sim {\rm{{\cal C}{\cal N}}}\left( {0,\kappa _{SR}^2{{ P}_S}{{\left| {{h_{SR}}} \right|}^2}} \right)$ and ${\eta _2} \sim {\rm{{\cal C}{\cal N}}}\left( {0,\kappa _{LI}^2{{ P}_R}{{\left| {{h_{LI}}} \right|}^2}} \right)$ characterize the distortion noise, ${\kappa _{SR}}$ and ${\kappa _{LI}}$ are the levels of RHIs; ${n_0} \sim {\rm{{\cal C}{\cal N}}}\left( {0,{N_0}} \right)$ denotes the additive white Gaussian noise (AWGN). $R$ decodes and forwards the received signals to users at the same time. Thus, the received signals at $D_f$ and $D_n$ are expressed as
\begin{equation}
\label{yrdf}
{y_{RI}} \!\!=\! {h\!_{RI}}\!\left(\!\! {\sqrt {{b_1}{P_R}} {x_1} \!\!+\! \!\sqrt {{b_2}{P_R}} {x_2} \!+\! {\eta _I}} \! \right) \!+ \!{n_0},I \!\in\! \left\{\! {{D\!_f},\!{D\!_n}} \!\right\},
\end{equation}
where ${\eta _I} \sim {\rm{{\cal C}{\cal N}}}\left( {0,\kappa _{RI}^2{P_R}{{\left| {{h_{RI}}} \right|}^2}} \right)$ characterizes the distortion noise; ${h_{RI}}$ is the channel coefficient of $R \to I$; ${\kappa _{RI}}$ is the level of RHIs; $b_1$ and $b_2$ are the power allocation coefficients of relay transmitted signals with ${b_1} + {b_2} = 1$ and ${b_1} > {b_2}$.

Leveraging NOMA protocol, $R$ first decodes $D_f$'s signal $x_1$ by invoking SIC, then the signal $x_2$ will be decoded. Therefore, the received signal-to-interference-plus-noise ratios (SINRs) of signal $x_1$ and $x_2$ at $R$ are expressed as
\begin{equation}
\label{srx1}
\gamma _{{x_1}}^{SR} = \frac{{{a_1}{\rho _{SR}}\gamma }}{{\left( {{a_2} + \kappa _{SR}^2} \right){\rho _{SR}}\gamma  + {\rho _{LI}}{\varpi ^2}\gamma '\left( {1 + \kappa _{SR}^2} \right) + 1}},
\end{equation}
\begin{equation}
\label{srx2}
\gamma _{{x_2}}^{SR} = \frac{{{a_2}{\rho _{SR}}\gamma }}{{{\rho _{SR}}\kappa _{SR}^2\gamma \! +\! {\rho _{LI}}{\varpi ^2}\gamma '\left( {1 \!+\! \kappa _{SR}^2} \right) \!+\! {a_1}{g_{SR}}\gamma \! +\! 1}},
\end{equation}
where $\gamma  = {{ P}_S}/{N_0}$ and $\gamma ' = {{ P}_R}/{N_0}$ are the transmit SNR at $S$ and $R$, respectively; ${\rho _{SR}} = {\left| {{h_{SR}}} \right|^2}$ and ${\rho _{LI}} = {\left| {{h_{LI}}} \right|^2}$ are the channel gains; ${g_{SR}} \sim {\cal C}{\cal N}\left( {0,\varepsilon {\rho _{SR}}} \right)$ and $\varepsilon  \in \left[ {0,1} \right)$\footnote{It is noted that when $\varepsilon=0$, the effects of the $x_1$ on the near user are absent and the system performs perfect SIC, while $\varepsilon$ cannot reach 1 which is due to the fact that $\varepsilon=1$ implies signals cannot decode successfully at the users.} is the percentage of residual signal $x_1$ caused by ipSIC which follows Gaussian distribution \cite{lixingwcl19}. Then, the decoded signals are forwarded to $D_f$ and $D_n$ by $R$. Thus, the received SINR at $D_f$ is expressed as
\begin{equation}
\label{rdfx1}
\gamma _{{x_1}}^{R{D_f}} = \frac{{{b_1}{\rho _{R{D_f}}}\gamma '}}{{{b_2}{\rho _{R{D_f}}}\gamma ' + {\rho _{R{D_f}}}\kappa _{R{D_f}}^2\gamma ' + 1}},
\end{equation}
where ${\rho _{R{D_f}}} = {\left| {{h_{R{D_f}}}} \right|^2}$. Similarly, the SINRs for $D_n$ to decode the desired signal and $D_f$'s signal are expressed as
\begin{equation}
\label{rdnx2}
\gamma _{{x_2}}^{R{D_n}} = \frac{{{b_2}{\rho _{R{D_n}}}\gamma '}}{{{\rho _{R{D_n}}}\kappa _{R{D_n}}^2\gamma ' + {b_1}{g_{R{D_n}}}\gamma ' + 1}},
\end{equation}
\begin{equation}
\label{rdnx1}
\gamma _{{x_1}}^{R{D_n}} = \frac{{{b_1}{\rho _{R{D_n}}}\gamma '}}{{{b_2}{\rho _{R{D_n}}}\gamma ' + {\rho _{R{D_n}}}\kappa _{R{D_n}}^2\gamma ' + 1}},
\end{equation}
where ${\rho _{R{D_n}}} = {\left| {{h_{R{D_n}}}} \right|^2}$ and ${g_{RD_n}} \sim {\cal C}{\cal N}\left( {0,\varepsilon {\rho _{RD_n}}} \right)$.
\subsection{Fading Channel}
The probability density function (PDF) of the Rician channel gain ${\rho _i}{\rm{ }}\left( {i = SR,{\rm{ }}LI,{\rm{ }}R{D_f},{\rm{ }}R{D_n}} \right)$ can be expressed as
\begin{equation}
\label{pdfpdf}
{f_{{\rho _i}}}\!\left( x \right) = \frac{{\left(\! {K\! + \!1}\! \right){e^{ - K}}}}{{{\lambda _i}}}{e^{ - \frac{{\left( \!{K\! + \!1}\! \right)x}}{{{\lambda _i}}}}}{I_0}\left( {2\sqrt {\frac{{K\left( {K + 1} \right)x}}{{{\lambda _i}}}} } \right),
\end{equation}
where ${\lambda _i}$ is the mean value of ${\rho _i}$, $K$ is the Rician $K$-factor defined as the ratio of the power of the line-of-sight (LOS) component to the separate components and ${I_0}\left(  \cdot  \right)$ denotes the zero-th order modified Bessel function of the first kind. Using \cite{Suraweera2009}, the PDF of the Rician channels can be rewritten as
\begin{equation}
\label{pdf}
{f_{{\rho _i}}}\!\left( x \right){\rm{ = }}\frac{{\left( \!{K \!+\! 1}\! \right){e^{ - K}}}}{{{\lambda _i}}}\sum\limits_{l = 0}^\infty  {\frac{{{x^l}}}{{{{\left( {l!} \right)}^2}}}} {\left[ {\frac{{K\!\left(\! {K \!+\! 1}\! \right)}}{{{\lambda _i}}}} \right]^l}{e^{ - \frac{{K + 1}}{{{\lambda _i}}}x}}.
\end{equation}

\section{Performance Analysis}
In this section, the analytical expressions of OP for $D_f$ and $D_n$ are derived and we also derive the ESR in high SNRs to reflect the performance of the considered system.
\subsection{OP Analysis}
For $D_f$, the outage event will not occur when relay can successfully decodes $x_1$ and $x_2$, and $D_f$ can successfully decode signal $x_1$. Thus the OP of $D_f$ can be expressed as
\begin{equation}
\label{poutdf}
{\rm P}_{out}^{{D_f}} =  \Pr \left( {\min \left( {\frac{{\gamma _{{x_1}}^{SR}}}{{{\gamma _{thf}}}},\frac{{\gamma _{{x_1}}^{R{D_f}}}}{{{\gamma _{thf}}}},\frac{{\gamma _{{x_2}}^{SR}}}{{{\gamma _{thn}}}}} \right) < 1} \right),
\end{equation}
where ${\gamma _{thf}}$ and ${\gamma _{thn}}$ are the target threshold at $D_f$ and $D_n$, respectively. The analytical expression of the OP is provided in the following theorem.
\begin{theorem}
The analytical expression of the OP for $D_f$ can be expressed as
\begin{align} \nonumber
&{\rm P}_{out}^{{D_f}} = 1 - \sum\limits_{{l_1} = 0}^\infty  {\sum\limits_{{l_2} = 0}^\infty  {\sum\limits_{{l_3} = 0}^\infty  {\sum\limits_{{m_1} = 0}^{{l_2}} {\sum\limits_{{m_2} = 0}^{{l_3}} {\sum\limits_{u = 0}^{{m_1}} {\left( {\begin{array}{*{20}{c}}
{{m_1}}\\
u
\end{array}} \right)} } } } } }\frac{{{K^{{l_1} + {l_2} + {l_3}}}}}{{{m_1}!{m_2}!}} \\\nonumber
 &\times \left( {u + {l_1}} \right)!{\chi ^u}\lambda _{LI}^u{\xi ^{{m_1} - u}}\lambda _{SR}^{{l_1} + u - {m_1} + 1}{\left( {{\lambda _{SR}} + {\lambda _{LI}}\chi } \right)^{ - \left( {{l_1} + u + 1} \right)}}\\\label{poutDF}
& \times \frac{{{{\left( \!{K \!+\! 1}\! \right)}^{{m_1}+m_2 - u}}}}{{{{\left( {{l_1}!} \right)}^2}{l_2}!{l_3}!}}{\left(\! {\frac{{\psi }}{{{\lambda _{R{D_f}}}}}}\! \right)^{{m_2}}}{e^{ - \frac{{\left(\! {K \!+\! 1}\! \right)\xi }}{{{\lambda _{SR}}}} - \frac{{\left(\! {K\! +\! 1}\! \right)\psi }}{{{\lambda _{R{D_f}}}}} - 3K}},
\end{align}
\end{theorem}
\noindent where ${\xi _1} = {\gamma _{thf}}/\left({a_1}\gamma  - \left( {{a_2} + \kappa _{SR}^2} \right)\gamma {\gamma _{thf}}\right)$, ${\xi _2} = {\gamma _{thn}}/\left( {{a_2}\gamma  - \left( {{a_1}\varepsilon  + \kappa _{SR}^2} \right)\gamma {\gamma _{thn}}} \right)$, $\xi  = \max \left( {{\xi _1},{\xi _2}} \right)$, $\chi  = {\varpi ^2}\gamma '\left( {\kappa _{SR}^2 + 1} \right)\xi $, $\psi  = {\gamma _{thf}}/\left( {{b_1}\gamma ' - \left( {{b_2} + \kappa _{R{D_f}}^2} \right)\gamma '{\gamma _{thf}}} \right)$.
\begin{proof}
Substituting (\ref{srx1}) and (\ref{rdfx1}) into (\ref{poutdf}), the OP of $D_f$ can be further expressed as
\begin{align} \nonumber
{\rm P}_{out}^{{D_f}} &= 1 - \Pr \left( {\gamma _{{x_1}}^{SR} > {\gamma _{thf}},\gamma _{{x_1}}^{R{D_f}} > {\gamma _{thf}}},\gamma _{{x_2}}^{SR} > {\gamma _{thn}} \right)\\ \label{proofdf}
 &= 1 \!- \!\left(\!1\!-\!{F_{{\rho _{R{D_f}}}}}\!\left( \psi  \right)\!\right)\!\!\int_0^\infty \!\!\! {{f_{{\rho _{LI}}}}\!\!\left( y \right)\!{F_{{\rho _{SR}}}}\!\left( {\chi y\! +\! \xi } \right)} dy,
\end{align}
substituting (\ref{pdf}) into (\ref{proofdf}), utilizing \cite[Eq. (3.478.1)]{Gradshteyn2007}, the result can be obtained after some mathematical operations.
\end{proof}
For $D_n$, outage events will not occur when both $R$ and $D_n$ decode $x_1$ and $x_2$ successfully, thus the OP of $D_n$ can be expressed as
\begin{align} \label{poutdn}
&{\rm P}_{out}^{{D_n}} = \Pr \left( {\min \left( {\frac{{\gamma _{{x_1}}^{SR}}}{{{\gamma _{thf}}}},\frac{{\gamma _{{x_2}}^{SR}}}{{{\gamma _{thn}}}}} \right) < 1} \right)\\\nonumber
 &+ \Pr \left( {\min \left( {\frac{{\gamma _{{x_1}}^{SR}}}{{{\gamma _{thf}}}},\frac{{\gamma _{{x_2}}^{SR}}}{{{\gamma _{thn}}}}} \right) \ge 1,\min \left( {\frac{{\gamma _{{x_1}}^{R{D_n}}}}{{{\gamma _{thf}}}},\frac{{\gamma _{{x_2}}^{R{D_n}}}}{{{\gamma _{thn}}}}} \right) < 1} \right),
\end{align}

The analytical expression of the OP will be given in the following theorem.
\begin{theorem}
The analytical expression of the OP of $D_n$ can be expressed as
\begin{align} \nonumber
&{\rm P}_{out}^{{D_n}}{\rm{ }}=\!\!1 \!\!-\!\! \!\sum\limits_{{l\!_4} = 0}^\infty \! {\sum\limits_{{l\!_5} = 0}^\infty  \!{\sum\limits_{{l\!_6} = 0}^\infty\! {\sum\limits_{{m\!_3} = 0}^{{l_5}}\! {\sum\limits_{{m\!_4} = 0}^{{l_6}} \! {\sum\limits_{t = 0}^{{m_3}} \!\!{\left( \!\!\!\!{\begin{array}{*{20}{c}}
{{m_3}}\\
t
\end{array}}\!\!\!\! \right)\!\frac{{{K^{{l_4}\! + \!{l_5} \!+ \!{l_6}}}\!{{\left(\! {K\!\! +\!\! 1}\! \right)}^{{m_3}\! - \!t}}}}{{{{\left( {{l_4}!} \right)}^2}{l_5}!{l_6}!{m_3}!{m_4}!}}} } } } } } \\ \nonumber
&\times \left( {t + {l_4}} \right)!{\chi  ^t}{\xi  ^{{m_3} - t}}\lambda _{LI}^t\lambda _{SR}^{{l_4} + t - {m_3} + 1}{\left( {{\lambda _{SR}} + {\lambda _{LI}}\chi } \right)^{ - \left( {{l_4} + t + 1} \right)}}\\ \label{Poutdf}
&\times {\left( {\frac{{\phi \left( {K + 1} \right)}}{{{\lambda _{R{D_n}}}}}} \right)^{{m_4}}}{e^{ - \left( {K + 1} \right)\left( {\frac{\chi  }{{{\lambda _{SR}}}} + \frac{\phi }{{{\lambda _{R{D_n}}}}}} \right) - K}},
\end{align}
\end{theorem}
\noindent where $\phi  = \max \left( {{\phi _1},{\phi _2}} \right)$, ${\phi _1} = {\gamma _{thf}}/\left( {{b_1}\gamma ' - \left( {{b_2} + \kappa _{R{D_n}}^2} \right)\gamma '{\gamma _{thf}}} \right)$, ${\phi _2} = {\gamma _{thn}}/\left( {{b_2}\gamma ' - \left( {\varepsilon {b_1} + \kappa _{R{D_n}}^2} \right)\gamma '{\gamma _{thn}}} \right)$.
\begin{proof}
Substituting (\ref{srx1}), (\ref{srx2}), (\ref{rdnx2}) and (\ref{rdnx1}) into (\ref{poutdn}), using the similar methodology of Theorem 1, the result can be obtained after some basic operations.
\end{proof}

\subsection{Diversity Order}
Diversity gain is a key metric to evaluate the effects of fading parameters on system performance at high SNRs. The diversity order of the users is defined as \cite{liuyuan2017}
\begin{align}
\label{order}
d =  - \mathop {\lim }\limits_{\Upsilon   \to \infty } \frac{{\log \left( {{\rm P}_{out}^h} \right)}}{{\log \Upsilon  }},
\end{align}
where ${{\rm P}_{out}^h}$ denotes the asymptotic OP, $\Upsilon$ is the transmit SNR.

To obtain the diversity order, we first carry out the asymptotic OPs of the users in the following corollary.

\begin{corollary}
Based on the OPs of two users, the asymptotic OPs at high SNRs for $D_f$ and $D_n$ are given as
\begin{align} \nonumber
{\rm P}_{out}^{f,h}&  =  1 - \left( {1 - {e^{ - K}}\frac{{\left( {K + 1} \right)\psi }}{{{\lambda _{R{D_f}}}}}} \right)\left[ {1 - \sum\limits_{{l_1}^\prime  = 0}^\infty  {\frac{{{K^{{l_1}^\prime }}{e^{ - 2K}}}}{{{\lambda _{SR}}{l_1}^\prime !}}} } \right.\\\label{poutfh}
&\left. { \times \left( {{\lambda _{LI}}\chi \left( {{l_1}^\prime  + 1} \right) + \left( {K + 1} \right)\xi } \right)} \right],
\end{align}

\begin{align} \nonumber
{\rm P}_{out}^{n,h} & =  1 - \left( {1 - {e^{ - K}}\frac{{\left( {K + 1} \right)\phi }}{{{\lambda _{R{D_n}}}}}} \right)\left[ {1 - \sum\limits_{{l_2}^\prime  = 0}^\infty  {\frac{{{K^{{l_2}^\prime }}{e^{ - 2K}}}}{{{\lambda _{SR}}{l_2}^\prime !}}} } \right.\\\label{poutnh}
&\left. { \times \left( {{\lambda _{LI}}\chi \left( {{l_2}^\prime  + 1} \right) + \left( {K + 1} \right)\xi } \right)} \right].
\end{align}
\end{corollary}

\begin{corollary}
Based on the asymptotic OPs of two users, the diversity orders of $D_f$ and $D_n$ in the ideal $\left({\kappa  = \varepsilon  = 0}\right)$ and non-ideal conditions $\left( {\kappa ,\varepsilon  \ne 0} \right)$ are given as
\begin{align}
d_f^{id} = d_n^{id} = d_f^{nid} = d_n^{nid} = 0.
\end{align}

\end{corollary}

\begin{remark}
The results show the effects of channel fading and non-ideal parameters on the outage performance intuitively. For non-ideal conditions, RHIs and ipSIC have detrimental effects on the outage performance of the considered system. Moreover, the OPs keep a constant value and the diversity orders of $D_f$ and $D_n$ are 0 at high SNRs due to LSI. In addition, it is worth mentioning that when ${\kappa  = \varepsilon  = 0}$, the considered system reduces to the ideal conditions.
\end{remark}

\subsection{ESR Analysis}
ESR is another metric to evaluate the system performance, which is defined as the sum rate of each user. The achievable rates of $D_f$ and $D_n$ are expressed as \cite{lixingwcl19}
\begin{equation}
\label{rf}
{R_f} = {\log _2}\left( {1 + \min \left[ {\gamma _{{x_1}}^{SR},\gamma _{{x_1}}^{R{D_f}}} \right]} \right),
\end{equation}
\begin{equation}
\label{rn}
{R_n} = \left\{ {\begin{array}{*{20}{c}}
{0,\;\;\;\;\;\;\;\;\;\;\;\;\;\;\;\;\;\;\;\;\;\;\;{\rm{  if }}{{\left| {{h_{R{D_n}}}} \right|}^2} < {{\left| {{h_{R{D_f}}}} \right|}^2}}\\
{{{\log }_2}\left( {1 + \min \left[ {\gamma _{{x_2}}^{SR},\gamma _{{x_2}}^{R{D_n}}} \right]} \right){\rm{,  otherwise}}}
\end{array}} \right..
\end{equation}
Since the channels ${h_{R{D_n}}}$ and ${h_{R{D_f}}}$ are independent random variables, we assume that $\Pr \left( {{{\left| {{h_{R{D_n}}}} \right|}^2} < {{\left| {{h_{R{D_f}}}} \right|}^2}} \right) = \Pr \left( {{{\left| {{h_{R{D_n}}}} \right|}^2} \ge {{\left| {{h_{R{D_f}}}} \right|}^2}} \right) = \frac{1}{2}$. Based on this, the ergodic rates (ERs) of $D_f$ and $D_n$ are given by
\begin{align}
\label{ravef}
R_{ave}^f = \mathbb{E}\left[ {{{\log }_2}\left( {1 + \min \left[ {\gamma _{{x_1}}^{SR},\gamma _{{x_1}}^{R{D_f}}} \right]} \right)} \right],
\end{align}
\begin{align}
\label{raven}
R_{ave}^n = \frac{1}{2}\mathbb{E}\left[ {{{\log }_2}\left( {1 + \min \left[ {\gamma _{{x_2}}^{SR},\gamma _{{x_2}}^{R{D_n}}} \right]} \right)} \right].
\end{align}
The  ERs for $D_f$ and $D_n$ at high SNRs are expressed as
\begin{align}\nonumber
R_{ave}^{f\_h} &= {\log _2}\left( {1 + \min \left[ {\frac{{{b_1}}}{{{b_2} + \kappa _{R{D_f}}^2}}} \right.} \right.,\\ \label{ravefh}
&\;\;\;\;\;\;\;\left. {\left. {\frac{{{a_1}{\Psi_{SR}}}}{{\left( {{a_2} + \kappa _{SR}^2} \right){\Psi _{SR}} + {\Psi _{LI}}{\varpi ^2}\left( {1 + \kappa _{SR}^2} \right)}}} \right]} \right),
\end{align}
\begin{align} \nonumber
R_{ave}^{n\_h} &= \frac{1}{2}{\log _2}\left( {1 + \min \left[ {\frac{{{b_2}}}{{\kappa _{R{D_n}}^2 + \varepsilon {b_1}}},} \right.} \right.\\ \label{ravedh}
&\;\;\;\;\;\;\;\left. {\left. {\frac{{{a_2}{\Psi_{SR}}}}{{{\Psi_{SR}}\left( {\kappa _{SR}^2 + \varepsilon {a_1}} \right) + {\Psi_{LI}}{\varpi ^2}\left( {1 + \kappa _{SR}^2} \right)}}} \right]} \right),
\end{align}
where ${\Psi_{SR}} ={\lambda _{SR}}{e^{ - K}}\sum\nolimits_{{l_1} = 0}^\infty  {\left( {{l_1} + 1} \right){K^{{l_1}}}/\left( {\left( {K + 1} \right){l_1}!} \right)} $, ${\Psi_{LI}} ={\lambda _{LI}}{e^{ - K}}\sum\nolimits_{{l_2} = 0}^\infty  {\left( {{l_2} + 1} \right){K^{{l_2}}}/\left( {\left( {K + 1} \right){l_2}!} \right)}$.
Thus, the ESR can be obtained as follows
\begin{equation}
\label{ravesum}
R_{ave}^{sum} = R_{ave}^f + R_{ave}^n; \;\;\;\;R_{ave}^{sum\_h} = R_{ave}^{f\_h} + R_{ave}^{n\_h}.
\end{equation}

\section{Numerical Results}
In this section, the correctness of the theoretical results is verified by Monte Carlo simulations. Unless other stated, the parameters of the Monte Carlo simulation are set as ${a_1}=b_1 = 0.7,{\rm{ }}{a_2}=b_2 = 0.3,{\rm{ }}\varepsilon  = 0.01,{\rm{ }}{\lambda _{SR}} = {\lambda _{R{D_f}}} = 8,{\rm{ }}{\lambda _{LI}} = 0.5,{\rm{ }}{\lambda _{R{D_n}}} = 1,{\rm{ }}{N_0} = 1$, $\left\{ {{\gamma _{thf}},{\gamma _{thn}}} \right\} = \left\{ {1,3} \right\}$ and $\left\{ {{\gamma _{thf}},{\gamma _{thn}}} \right\} = \left\{ {0.5,1.5} \right\}$. Moreover, hardware impairment levels are set: ${\kappa _{SR}} = {\kappa _{R{D_f}}} = {\kappa _{S{D_n}}} = 0.05$ \cite{lixingwsj}.
\begin{figure}[!tp]
\setlength{\abovecaptionskip}{0pt}
\centering
\includegraphics [width=3.0in]{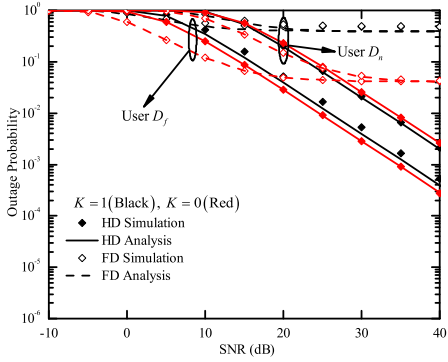}
\caption{OP of users vs. transmit SNR}
\label{fig1}
\end{figure}

Fig. 2 depicts the OPs of $D_f$ and $D_n$ versus transmit SNR. The curves of OP are given in the case of $K = 0$ and $K=1$. When $K=0$, the channels are reduced to Rayleigh channel. As can be seen from Fig. 2, for the HD relay systems, the system OP decreases with the increase of SNR. However, in the FD mode, there will be an error floor for the OP of the considered system, which fully indicates that due to the existence of LSI, the users' OP finally tends to be fixed. It can also be seen that whether for HD or FD mode, there is an intersection between the OPs of $D_f$ and $D_n$, which is due to the fact that LSI is not the major factor affecting the performance of the system when the system is in low SNR region.

\begin{figure}[htb]
\centering
\subfigure[]
{\begin{minipage}[t]{0.4\linewidth}
\centering
\includegraphics[width= 1.8in, height=1.8in]{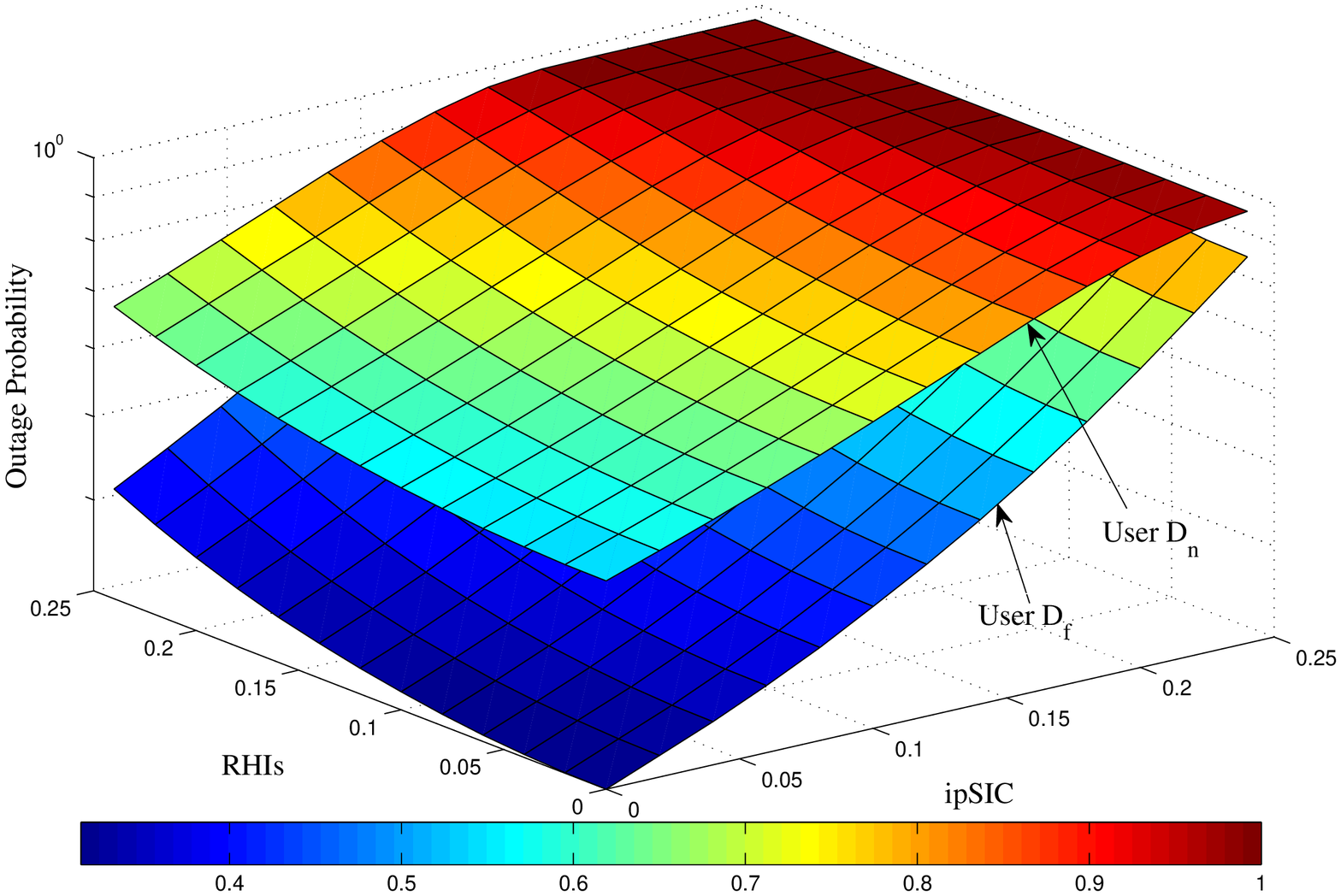}
\end{minipage}
}
\centering
\subfigure[]
{\begin{minipage}[t]{0.48\linewidth}
\centering
\includegraphics[width= 1.8in, height=1.8in]{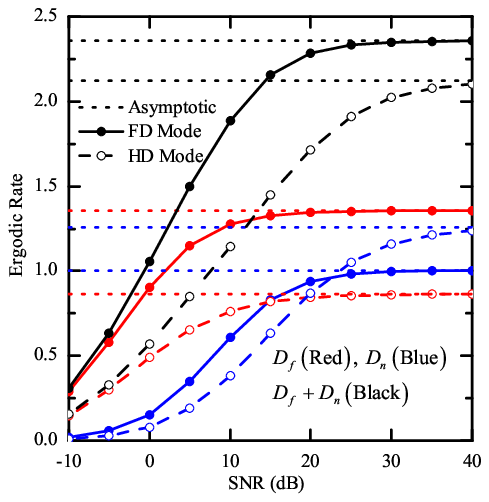}
\end{minipage}
}
\caption{(a) OP vs. RHIs and ipSIC; (b) ERs vs. transmit SNR}
\end{figure}


Fig. 3(a) plots the OPs of $D_f$ and $D_n$ versus ipSIC and RHIs in the FD mode with ${\rm{SNR = 10dB}}$. It can be observed from Fig. 3(a) that the OP of $D_f$ is lower than that of $D_n$, which is due to the higher power allocated to $D_f$. In addition, for both $D_f$ and $D_n$, the fluctuation for the OP of ipSIC is more obvious than that of RHIs, which shows that the outage performance of the users is more dependent on the ability of SIC. Furthermore, we can also note that the OP of $D_f$ increases drastic drastically than that of $D_n$ regardless of changing the RHIs or ipSIC which is due to the near users eliminate part of interference caused by far users.

Fig. 3(b) illustrates the ERs of the users versus transmit SNR with $K=1$. From Fig. 3(b), we can see that the ER of $D_f$ in FD mode is always higher than that of the HD mode, while for $D_n$, the ER in the FD mode is higher only in the case of low SNR region, which indicates that due to the influence of ipSIC, the LSI has a serious negative impact on $D_n$. On the other hand, the ergodic performance in the FD mode always outperforms the HD mode, this phenomenon indicates that although the considered system of FD mode sacrifices the OP, it significantly improves the ER of the system compares with the HD mode. It is worth noting that with the increase of SNR, the ESR of the system finally tends to a fixed value, which shows that the performance of the system cannot be improved only by simply improving the SNR.
\section{Conclusion}
We studied the performance of the FD DF cooperative NOMA system over Rician channels in the presence of ipSIC and RHIs. The analytical expressions for the OP of the users are derived. Furthermore, we also explored the ergodic performance of the considered system. The results show us that the LSI caused by relay has a great impact on the considered system. It can also be obtained from the simulation results that the extent of the negative impact of the ipSIC on the near user is serious than RHIs. Furthermore, it can also be concluded that compared with the HD cooperative NOMA system, the proposed FD cooperative NOMA system can greatly improve the ER performance of the users.

\bibliographystyle{IEEEtran}
\bibliography{myreference}

\end{document}